\documentstyle[12pt,openbib,psfig]{article}
\newcommand{\f}{\frac}

\newcommand{\be}{\beta}

\newcommand{\g}{\gamma}

\newcommand{\ep}{\epsilon}

\def\be{\begin{equation}}
\def\ee{\end{equation}}

\begin{document}

\title{\bf Stability of strange stars (SS) derived from
a realistic equation of state.}

\author{ Monika Sinha $^{1,~2,~3}$,  Jishnu Dey $^{2,~ 4~ \dagger}$,
Mira Dey $^{1,~ 4 ~\dagger}$,\\ Subharthi Ray $^{5}$ and
Siddhartha Bhowmick, $^{6}$ }

\maketitle

\begin{abstract}

  { A realistic EOS (equation of state) leads to strange stars (ReSS)
which are compact in the mass radius plot, close to the
Schwarzchild limiting line \cite{d98}. Many of the observed stars
fit in with this kind of compactness, irrespective of whether they
are X-ray pulsars, bursters or soft $\gamma$ repeaters or even
radio pulsars. We point out that a change in the radius of a
star can be small or large, when its mass is increasing and this
depends on the position of a particular star on the mass radius
curve. We carry out a stability analysis against radial
oscillations and compare with the EOS of other SS models. We find
that the ReSS is stable and an M-R region can be identified to that
effect.}

\end{abstract}
\vskip .2cm

\noindent(PACS Numbers: 95.30.Cq -- 97.10.q -- 97.10.Cv -- 97.10.Pg -- 97.10.Sj -- 97.60.Sm -- 12.39.x -- 51.30.i.)


\vskip .2cm keywords: compact stars~--~realistic strange
stars~--~dense matter~--~elementary particles ~--~ equation of
state

\vskip .2cm 

\noindent $^1$ Dept. of Physics, Presidency
College, 86/1 College Street, Kolkata 700 073, India\\ $^2$ Azad
Physics Centre, Dept. of Physics, Maulana Azad College, 8  Rafi
Ahmed Kidwai Road, Kolkata 700 013, India and IUCAA, Pune, India\\
$^3$ CSIR-NET fellow, Govt. of India.\\ $^4$ Associate, IUCAA,
Pune, India \\ $^5$ FAPERJ Fellow, Govt. of Brazil, Instituto de
Fisica, Universit\'e Federal Fluminense, Niteroi, RJ, Brasil\\
$^6$ Department of Physics, Barasat Govt. College, Barasat, North
24 Parganas, W. Bengal, India and IUCAA, Pune, India\\ $\dagger$
permanent address; 1/10 Prince Golam Md. Road, Kolkata 700
026,India;\\ e-mail : deyjm@giascl01.vsnl.net.in.\\ $^*$ Work
supported  in part  by DST grant no. SP/S2/K21/01, Govt. of India.
\newpage
\section{Introduction}
    Recently there has been some excitement about the possibility
that some compact stars are made from unusual forms of matter
\cite{jjd02}. Further, from an analysis of over 1 million seismic
data reports sent to the U.S. Geological Survey in the years
1990-93, which were not associated with traditional epicentral
sources, Anderson et al. (2002) claim to have found two events
that `have the properties predicted for the passage of a strange
quark nugget' through the earth \cite{anderson}. Approximately
8000 separate seismic stations around the world are included in
the database. If confirmed this would be a discovery similar to
the detection of gamma ray bursts .

\begin{figure}[htbp]
\centerline{\psfig{file=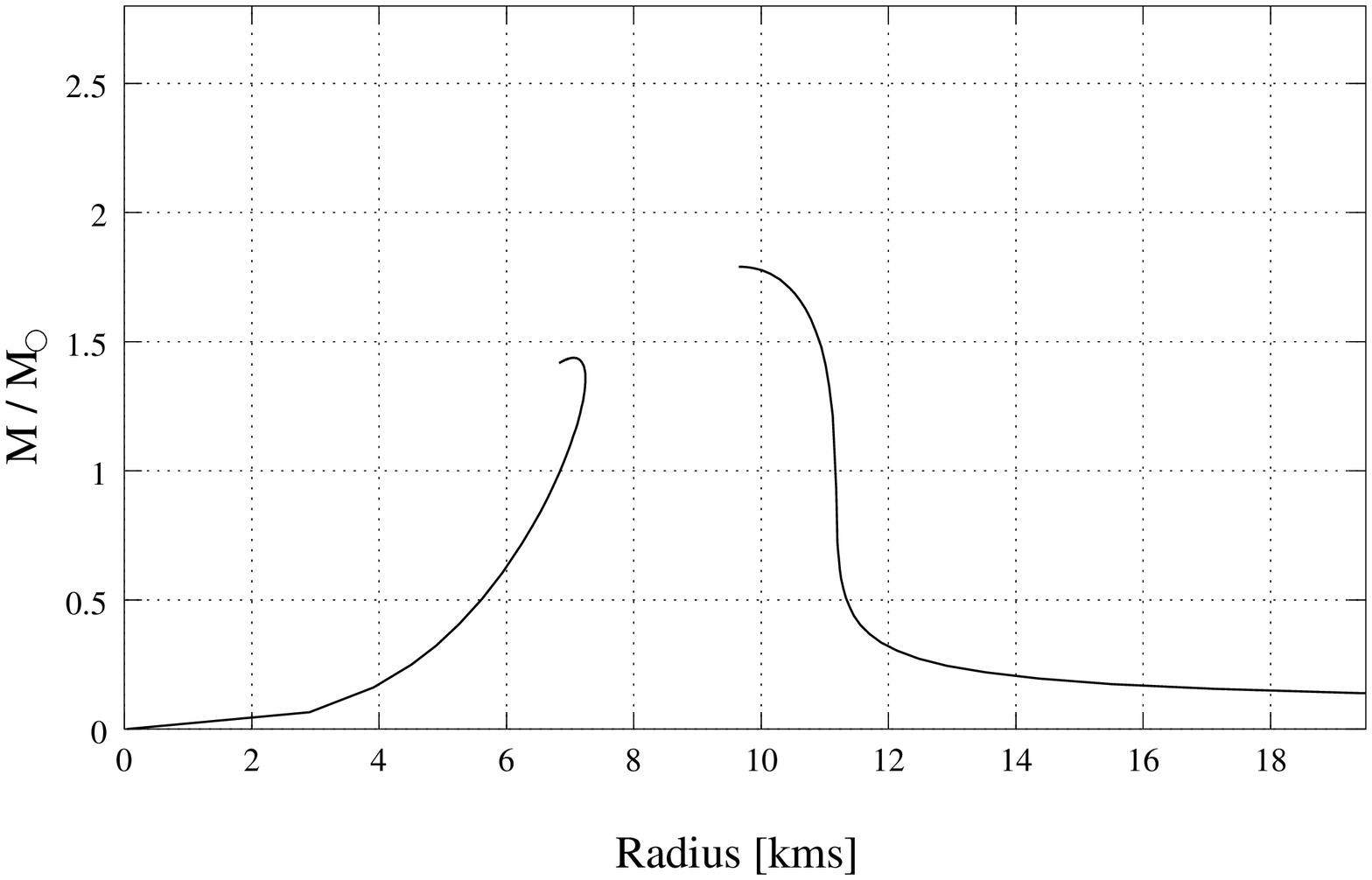,width=10cm}}
\caption{\footnotesize{The  mass and radius of stable stars with
the strange star EOS (left curve) and neutron star EOS (right
curve), which are solutions of the Tolman-Oppenheimer-Volkoff
(TOV) equations of general relativity. Note that self sustained
strange star systems can have small masses and radii, whereas
neutron stars have large radii for smaller masses since they are
bound by gravitation. }} \label{MR}
\end{figure}

    The best observational evidence for the existence of quark stars
seems to come from some compact objects like the X-Ray burst
sources SAX~J1808.4$-$3658 (the SAX in short) and 4U~1728$-$34,
the X-ray pulsar Her X-1 and the super burster 4U~1820$-$30. Among
these the first is the most stable pulsating X-ray source known to
man as of now. This star is claimed to be an ReSS with mass
\cite{Li99a} $\sim 1.3 ~M_\odot~$ and a radius of about 7 km. The
mass of 4U~1728$-$34 is claimed to be less than 1.1 $M_\odot$ in
Li et al. \cite{Li99b}, which places it much lower in the M-R plot
(Fig.\ref{MR}). So it could be still gaining mass and shift to
another stable point on the M-R graph. Thus in the model proposed
in \cite{d98} there is a possible answer to the question posed by
Franco \cite{fr01}: why are the pulsations of SAX not attenuated,
as they are in the 4U~1728$-$34 ?

Fig.\ref{MR} presents M-R relations for neutron as well as strange
stars. The current phenomenology of compact objects could be
interpreted to indicate that the mass of a star increases due to
accretion so that the radius of a star changes from one point of
the stable M-R curve to another. For neutron stars, exemplified by
the EOS BBB {\footnote {this is one of the set calculated by
Baldo, Bombaci and Burgeo \cite{bbb97aa} using a realistic nuclear
equation of state.}} - the curve on the right in Fig.\ref{MR} -  a
smaller mass would imply a larger radius for the star. If the mass
of the star increases it should contract. Therefore, expansion due
to an increase of mass, subsequent to accretion, is an unstable
process for a neutron star.  The expected behaviour for SS is
directly opposite to that of neutron stars as Fig.\ref{MR} shows
and therefore may support the claim that some compact stars are
ReSS.

    Coupled to our claim are various other evidences for the existence
of ReSS, such as explaining the compact M-R relations of the two
candidates given in \cite{d98} (namely, the Her X-1 and the
4U~1820$-$30), as also the possible explanation of two kHz
quasiperiodic oscillations in the 4U 1728~-~34 \cite{Li99b}.

    Recently we found that the matter of an ReSS may have diquarks on
the surface which could account for the delayed emission of huge
amounts of energy, after a thermonuclear catastrophe. This could
be a possible scenario for superbursts \cite{sinha}.

    Some of the stars like the SAX~J1808.8$-$3658  or the PSR~1937$+$21
are fast rotors. ReSS have the possibility of withstanding high
rotations which neutron stars or even bag SS cannot sustain.
The maximum frequencies for the two EOS of D98 are 2.6 and 2.8 kHz
respectively when they are on the mass shed limit (supramassive
model) and 1.8 kHz and 2 kHz when they are in the normal
evolutionary sequence as shown in Gondek-Rosi\'nska et al.
\cite{gbzgrdd00aa}.  In the present paper we further show that the
ReSS are not only stable under fast rotation but also against
radial oscillations.

The strange matter hypothesis has been used to postulate a
scenario whereby a neutron star collapses to a quark star,
simultaneously with a gamma ray burst - the so called phenomenon
of a quark nova \cite{qnova}.

In the cosmic separation of phase scenario of Witten \cite{wit} SS
are created along with baryons in a hot environment during the
expansion of the early universe. The problem was investigated for
ReSS \cite{r2000} and it was found that they are formed at a
temperature $T~\sim~70$ $MeV$. Since it is self-sustaining, the
system expands as the Universe cools, the ReSS expands like the
Universe itself with cooling \cite{r2000}. The calculation also
suggests that the shift in entropy due the change from normal to
strange matter is not very large, indicating that the phase
transition is relatively smooth.

    We note that in a series of early papers van Paradijs
(\cite{p78n}) had noted that (i) the X-rays for bursters originate
from stars with radii around 7 kms, assuming a canonical mass of
1.4 $M_\odot$ for them and (ii) if one assumes a lower mass  the
estimated radii also becomes lower, which fits the M-R relation
for the ReSS (Fig. \ref{MR})

   Because of the extremely strong electric field stretching outside
the stellar surface within $\sim10 ^{-10}$ cm \cite {a91npb} Usov
\cite{u1} suggested that at a finite surface temperature ($5\times
10^8~K$) of an SS one expects to have the creation of $e^+ e^-$
pairs on its surface. Such an effect could be an additional
observational signature of SS with nearly bare quark surfaces. The
strange matter equilibrium will not allow the pair production to
quench, contrary to the  comment in  \cite{m98} as pointed out in
\cite{u3}. Usov claimed that the chemical potential of the
electrons at the surface is large compared to their mass, being
around 20 MeV and the mean velocity of the electrons is very high,
so that the electric field will be restored very quickly. In our
calculations we specifically find this to be $\sim$29 MeV, thus
strongly supporting the conclusions of Usov.

     Usov has further claimed that from the nature of the two
step process, viz. $e^+ e^-$ pair production and subsequent $\g$
emission, the  soft $\g$ repeaters are indeed very young SS
\cite{u4} and their genesis may be due to the impact of comet like
objects with these stars \cite{u5}.

    We must also add that radio pulsars may be SS as was suggested
recently by Xu et al \cite{xqz99apj} for the PSR 0943+10 and all
other drifting pulsars. Further, Kapoor and Shukre \cite{ks} used
a remarkably precise observational relation for pulsar core
component widths of radio pulsars to get stringent limits on
pulsar radii, strongly indicating that some pulsars are strange
stars. This is achieved by including general relativistic effects
due to the pulsar mass on the size of the emission region needed
to explain the observed pulse widths. A recent paper supporting
their ideas is Xu, Xu and Wu \cite{xxw} for PSR 1937$+$21,the
fastest known radio pulsar.

    The calculations for cold ReSS by Dey et al. \cite{d98} enables us to
draw conclusions about chiral symmetry restoration (CSR) in QCD
when the EOS is used to get ReSS fitting definite M-R relations
\cite{d98, Li99a, Li99b}. The empirical M-R relations were derived
from astrophysical observations like luminosity variation and some
properties of quasiperiodic oscillations from compact stars. The
density dependence of the strong coupling constant  can be deduced
from the CSR described above \cite{srdd00mpla}.

    Recently Glendenning \cite{g00prl} has argued that the SAX could
be explained as a neutron star rather than a bare SS, not with any
of the existing known EOS, but with a hypothetical one, satisfying
however, the well-accepted restrictions based on general physical
principles and having a core density about 26 $\rho_0$. Of course,
such high density cores imply hybrid strange stars, subject to
Glendenning's assumption that such stars can exist with matching
EOS for two phases. There is the further constraint that if the
most compact hybrid star has a given mass, all lighter stars must
be larger. It was found in Li et al. \cite{Li99b} that the star
4U~1728$-$34 may have a mass less than that of the SAX and yet
have a smaller radius. Another serious difference is that the EOS
of D98, using the formalism of large $N_{\rm c}$ approximation,
indeed shows a bound state in the sense of having a minimum at about
$4.8~n_0$, whereas in Glendenning \cite{g00prl} one of the
assumptions is that strange matter has no bound state.

\section{ ReSS model - the  EOS }

    The model solves the relativistic mean field equation with a
realistic qq interaction,  the quark masses decreasing with
density and restoring to their current mass values at high enough
density. Moreover, the bare, confining qq interaction is expected
to get screened in the medium. The inverse Debye screening length
($D^{-1}$) is an increasing function of density. With these two
conditions, namely the tendencies towards deconfinement and chiral
symmetry restoration with density, we set out  to obtain the
required EOS for beta-equilibrated, charge neutral strange quark
matter (SQM)  containing uds quarks.

The plausibility condition for SQM to be preferred to ordinary matter is,
\be
E_{ud}/A~ > ~ E_{Fe^{56}}/A ~ > ~ E_{uds}/A \label{plausi} \ee

\begin{figure}[htbp]
\centerline{\psfig{file=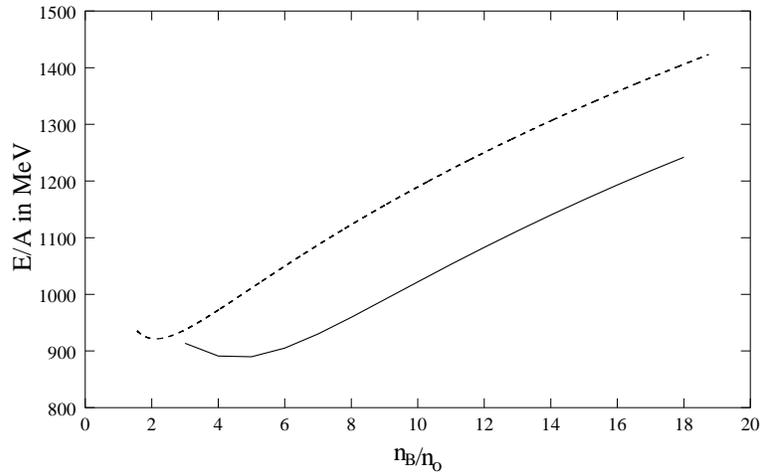,width=10cm}}

\caption{\footnotesize{The eos1 in d98 shows that there is a
stable point in the strange matter which has energy per baryon
less than that of $Fe^{56}$ by more than 40 $MeV$ (lower curve) .
The dashed curve shows a bag model EOS with B=72 MeV/$fm^3$ with
non-interacting massless u,d quarks and  $m_s$ = 150 MeV. (upper
curve)}} \label{efit}

\end{figure}

In Fig.\ref{efit} we have shown that the energy per baryon for our
equation of state (EOS1 of \cite{d98}) has a minimum at E/A =
888.8 $MeV$ compared to 930.4 of $Fe^{56}$. The pressure at this
point is zero and this marks the surface of the star as can be
seen when the well known TOV equation is solved. The curve clearly
shows that the system can fluctuate about this minimum - so that
the zero pressure point can vary. The EOS is parameterized to a
linear form in \cite{gbzgrdd00aa} as
\be
p=a(\ep - \ep_{0}), \label{linear} \ee
where $a$ and $\ep_{0}$ are two parameters, $a=0.463$ and
$\ep_{0}=1.15 \times 10^{15}~gm~cm^{-3}$.\\
For EOS3 they are $a=0.455$, $\ep_{0}=1.33 \times
10^{15}~gm~cm^{-3}$.
Sharma et al \cite{sharma} showed that the model possesses scaling
properties similar to the bag model.

    The bag model EOS, which satisfies the plausibility condition
Eq.(\ref{plausi}), must have a bag constant which cannot be higher
than $75 ~~MeV/fm^3$ when the strange quark mass is taken to be
$150 ~~MeV$. We have plotted this EOS as a dashed curve in
Fig.\ref{efit}. The minimum in E/A is seen to occur at a low
density, about $\sim  2n_0$ where $n_0$ is the normal nuclear
matter density of $0.17/fm^3$. This kind of density may have been
already reached in present day heavy ion collision experiments and
yet no clear signature of stable strange matter has been observed.
This problem does not occur with ReSS where the surface density is
about $\sim 5 ~n_0$. In addition to this, the bag does not fit the
M-R data for the strange star candidates as has been repeatedly
stated before \cite{d98, Li99a, Li99b}. The minimum of energy of
the bag model (dashed curve) is 921 $MeV$. The gain in energy in
strange matter, using the bag model, is only a few $MeV$ over
$Fe^{56}$, of the order of thermonuclear energy release. For the
realistic EOS it is much larger $\sim 40 ~MeV$.

    The unusually hard X-ray burster GRO~J1744$-$28 \cite{cdwl98s} or the
soft $\g$ repeaters \cite{cd98prl} require the strange star
models, since these stars require an energy release which is large
compared to thermonuclear energies. With the bag model that would
fit the requirement Eq.(\ref{plausi}), it is difficult to explain
the hard X-ray bursters or soft $\g$ repeaters since the energy
gain over $Fe^{56}$ is only $\sim 3$ $MeV$.

\section{Mass - Radius }

    Quasiperiodic oscillations in the 4U~1728$-$34, led to the idea
that it is an ReSS \cite{Li99b} with a mass 1.1$M_\odot$. The star
is 4.3 kpc away.

\begin{figure}[htbp]

\centerline{\psfig{figure=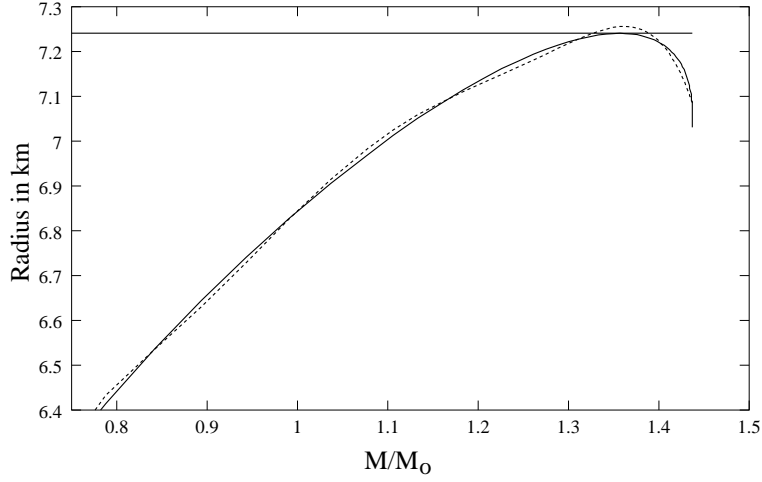,width=10cm}}
\caption{\footnotesize{The radius of a star plotted as a function
of its mass for EOS1. The horizontal line is the tangent drawn at
the maximum radius. The fitted polynomial is shown by a dashed
curve. }} \label{dRdM}
\end{figure}

    We elaborate on the point mentioned above, namely that if an SS gains
mass due to accretion its radius might not necessarily change. We
demonstrate this in Fig.\ref{dRdM} by the horizontal tangent line
drawn on the ReSS M-R relation. It is possible that the X-rays or
$\g$ rays emitted by a compact object is due to the conversion of
the accreting normal matter to strange matter \cite{boma}. The
energy gain by a baryon on conversion is nearly $\sim40$ $MeV$ or
so, making it a very favourable event. Thus a star of small mass
may become heavier with a consequent increase of the radius.
However, when it reaches the horizontal line $\frac{dR}{dM}$ is
zero and its radius does not increase for some time. We have
fitted the M-R curve to a polynomial,

\begin{equation}
R(M)~=~\sum_{i=1}^7~~a_i~\left(\f{M}{M_\odot}\right)^i \label{par}
\end{equation}

The parameters a$_i$ are given in Table \ref{param} for the
convenience of possible users. The parameters obtained from the
two EOS are given in Table \ref{eos}.

    It is found that some X-ray bursters like the 4U~1728$-$34 do not
show any radial expansion whereas others like the KS~1731$-$260
are usually observed only in bursts that exhibit photospheric
expansion \cite{muno}. We conjecture that the persistently bright
X-ray transient KS~1731$-$260 was a low mass star gaining mass and
thus a radius expansion was allowed by the TOV stability
criterion. After 11.5 years of activity this source was detected
in January 2001 with {{\it Chandra}} in quiescence with a
luminosity that is comparable to other normal X-ray emitting
compact stars \cite{chandra}. This could also be explained if the
star, in these decade of expansion, has reached the same stage as
the 4U~1728$-$34.

\section{ Radial oscillations of a relativistic star}

Thirty five years ago Chandrasekhar \cite{chan} investigated
these radial modes. They give information about the stability of
a stellar model. Recently such studies have been carried out for
many existing neutron star EOS \cite{kok}. Sharma et al
\cite{sharma} made a stability analysis of Eqn. (\ref{linear}) and
found that the EOS is stable against radial oscillations.
However, a more detailed  analysis is needed to restrict the M-R
relation. For completeness an outline of the scheme is given
below.

     The spherically symmetric metric is given by the line element
\be
ds^2 = -e^{2 \nu } dt^2 + e^{2 \mu
 } dr^2 + r^2 (d \theta^2 + \sin^2 \theta d\phi^2) .
\ee

Together with the energy-momentum tensor for a perfect fluid,
Einstein's field equations yield the TOV equation which can be
solved if we have an EOS, $p(n_B)$ and $\ep(n_B)$. Given the
central density $\ep_c$, we can arrive at an $M-R$ curve by
solving the TOV. Without disturbing the spherical symmetry of the
background we define $\delta r(r,t)$,  a time dependent radial
displacement of a fluid element located at the position $r$ in
the unperturbed model which assumes a harmonic time dependence, as

\be
\delta r(r,t)= u_{n}(r) e^{i\omega_{n} t}.
\ee

    The dynamical equation governing the stellar pulsation in its
$n$th normal mode ($n=0$, is the fundamental mode) has the
Sturm-Liouville's form ( for details, see \cite{mis} )

\be P(r) \frac{d^2 u_{n}(r)}{dr^2} + \frac{dP}{dr}
\frac{du_{n}}{dr} + \left[Q(r) + \omega_{n}^2 W(r)\right] u_{n}
(r) = 0, \ee where $u_{n}(r)$ and $\omega_{n}$ are the amplitude
and frequency of the $n$th normal mode, respectively. The
functions $P(r)$, $Q(r)$ and $W(r)$ are expressed in terms of the
equilibrium configuration of the star and are given by

\be P(r) = \frac{\Gamma p}{r^2}  e^{\mu+3\nu} \ee

\be
Q(r) = e^{\mu+3\nu}\left[\frac{(p')^2}{r^2(\ep+p)} -
\frac{4p'}{r^3} - \frac{8\pi}{r^2} (\ep + p) ~ p ~e^{2\mu}\right]
\ee

\be W = \frac{(\ep+p)}{r^2}~ e^{3\mu +\nu}, \ee
where the varying adiabatic index $\Gamma$ is  given by

\be \Gamma = \frac{(\ep+p)}{p}\frac{dp}{d\ep}, \ee $\ep$
and $p$ being the energy density and pressure of the unperturbed model,
respectively.
Eigenfrequencies can be obtained with the boundary conditions,
\begin{enumerate}
\item at the centre $r=0$, $\delta r= 0$ and
\item at the surface $\delta p =0$ leading to $\Gamma ~p~
u(r)'~=~0$,
\end{enumerate}

\begin{figure}[htbp]
\centerline{\psfig{figure=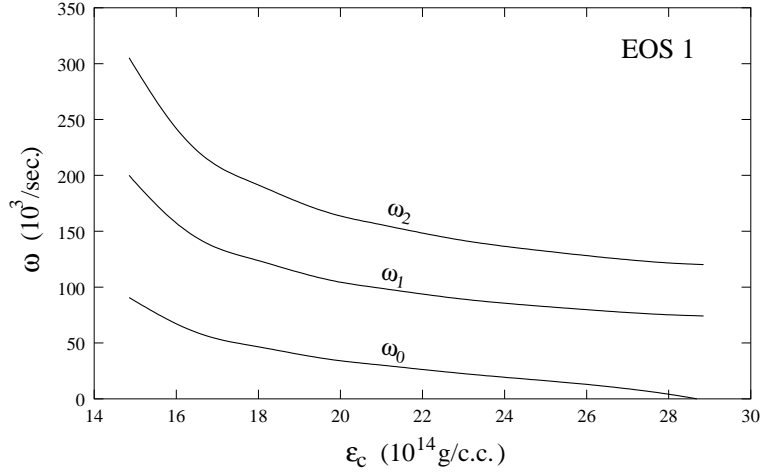,width=10cm,angle=0}}
\caption{\footnotesize{Angular frequency of three different modes
against central density for SS1.}} \label{freq}
\end{figure}

Since $\omega $ is real for $\omega ^2 > 0$, the solution is
oscillatory. However for $\omega ^2 < 0$, the angular frequency
$\omega$ is imaginary, which corresponds to an exponentially
growing solution. This means that for negative values of $\omega
^2 $ the radial oscillations are unstable. For a compact star the
fundamental mode $\omega_0$ becomes imaginary at some central
density $\ep_c$ less than the critical density $\ep_{critical}$
for which the total mass $M$ is a maximum. At $\ep_c=\ep_c^0$,
$\omega_0$ vanishes. All higher modes are zero at even higher
central densities. Therefore,  the star is unstable for central
densities greater than $\ep_c^0$. To illustrate, we plot the
eigenfrequencies  $\omega_n$ against $\ep_c$, the central density
in Fig. \ref{freq}. The fundamental frequency $\omega_0$ does
vanish at some $\ep_c^0$ while the higher modes remain nonzero.

Numerical values of  masses, radii, central densities and the
corresponding eigenfrequencies $\omega_0$, $\omega_1 $ and
$\omega_2$ are given in Tables \ref{eos1} and \ref{eos3} for our
EOS1 (SS1\cite{d98}) and EOS3(SS2\cite{d98}) respectively. The
corresponding linear fits described by Eqn.(\ref{linear}) for SS1
and SS2 are given in Table \ref{ss1} and \ref{ss2}. Tables
\ref{bag1} and \ref{bag2} are for the bag model EOS with different
parameters.

\section{ Discussions and summary}
In summary, evidence for the existence of strange stars have been
accumulating. In the present paper we review  possible
candidates  and suggest that the  properties peculiar to  some of
the compact stars can be explained if they are SS. In particular
we point out that mass accumulation due to accretion does not
lead to an increase in the radius for stars like the 4U 1728$-$34,
 claimed to be
low mass SS from accretion data \cite{Li99b}. If some compact objects
are proven to be ReSS then parameterizations of QCD chiral symmetry
restoration at high densities for quarks, the smallest particles
known, can be achieved with the help of data from some of the
heaviest objects in the Universe.

ReSS are stable against radial oscillations close to the maximum
attainable mass. For example, the EOS of SS1 sustains
gravitationally, $M_{max} \sim 1.4 M_{\odot}$, R=7 km with a
central number density $n_c ~\sim 16 n_0$. However, the
fundamental frequency of radial oscillations becomes zero at
around $n_c~9.5\sim n_0$,  destabilizing the star after M=1.36
$M_{\odot}$ with R= 7.24 km  (Table \ref{eos1}). It is still on
the $\frac{dM}{dR}>0$ region. Thus the maximum mass star which is
stable against radial oscillations has a  number density $\sim 9.5
n_0$ at the centre and $\sim 4.7 n_0$ at the surface.
Macroscopically,  upto this density small vibrations may be
sustained.

 A corresponding linear fit has the stable values M=1.34 $M_{\odot}$
 and R=7.35 km(Table \ref{ss1}). Radial oscillation is rather sensitive and the
 fit is better than 3$\%$(Tables \ref{eos1} and \ref{ss1}). The same pattern is seen for
 SS2 and the corresponding linear fit (Tables \ref{eos3} and \ref{ss2}). Thus we see
 that almost all of the M-R region is stable.

\begin{figure}[htbp]
\centerline{\psfig{figure=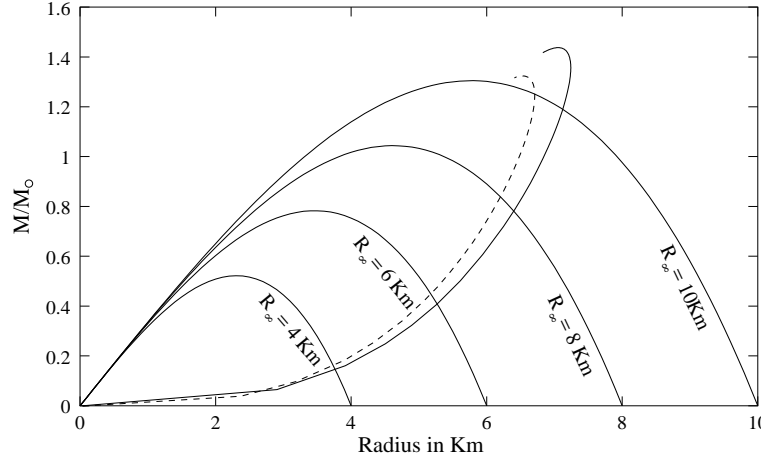,width=10cm}}
\caption{\footnotesize{Allowed M-R region with two EOSs, solid
curve for EOS1 and dashed curve for EOS3.}} \label{reg}
\end{figure}

There has been a recent controversy about the star
RXJ1856.$-$3754 $-$  whether it can be inferred to be a strange star
\cite{jjd02, pons}. Analysis of the observational data of this
``no pulsar'', 120 pc away, cool  star is not conclusive. As
noted by \cite {pr}  the stable portion of the $M-R$ region
shown in Fig. \ref{reg} can accommodate this star very easily.
This could be a possible candidate of our SS model provided its
mass and radius are established beyond controversy.

Acknowledgements:  The sad demise of our collaborator
Dr. Ranjan Ray was painful for us. We dedicate this paper to his memory.
MD and MS thank the RRI for hospitality and JD, MS and SB thank the IUCAA.





\begin{table}[htbp]

\caption{Parameters for the M-R curve in Fig.\ref {dRdM}}

\vskip 1cm

\begin{center}

\begin{tabular}{|c|c|c|c|c|c|c|c|}

\hline i&1&2&3&4&5&6&7\\ \hline
a$_i$&-343.7&2134&-5265&6805&-4884&1848&-288.4\\ \hline

\end{tabular}

\end{center}

\label{param}

\end{table}

\begin{table}[htbp]
\caption{Parameters from the two EOS} \vskip 1cm
\begin{center}
\begin{tabular}{|c|c|c|c|c|c|}
\hline EOS&$M_G/M_{\odot}$&R & $n_c/n_o$ & $\ep_c$ & $n_s/n_o$\\
&&(km)&&$(10^{14}g/c.c.)$&\\ \hline
SS1&1.437&7.055&13.669&46.85&4.586\\ \hline
SS2&1.325&6.5187&15.537&55.17&5.048\\ \hline
\end{tabular}

\end{center}

\label{eos}

\end{table}

\begin{table}[htbp]

\caption{Data for EOS1 (SS1)}

\vskip 1cm

\begin{center}

\begin{tabular}{|c|c|c|c|c|c|c|}
\hline

$\rho_c~10^{14}$&$n_c/n_0$&$M/M_{\odot}$&R&$\omega_0~10^3$&$\omega_1~10^3$&$\omega_2~10^3$\\
g/c.c.&&&km&/sec.&/sec.&/sec.\\

\hline

14.85&5.462&0.407&5.262&90.661&199.957&305.202\\
15.85&5.798&0.502&5.611&79.884&179.851&275.679\\
16.85&6.122&0.787&5.940&69.870&162.018&249.011\\
17.85&6.429&0.893&6.643&47.528&125.267&193.718\\
18.85&6.735&0.991&6.828&40.784&114.643&178.073\\
19.85&7.036&1.077&6.970&34.751&105.455&165.102\\
20.85&7.321&1.133&7.050&30.694&99.662&156.925\\
21.85&7.605&1.182&7.113&26.848&94.509&149.451\\
22.85&7.886&1.226&7.161&23.077&89.825&142.620\\
23.85&8.159&1.261&7.193&19.749&86.070&137.200\\
24.85&8.427&1.288&7.214&16.693&83.005&132.804\\
25.85&8.692&1.312&7.228&13.435&80.192&128.753\\
26.85&8.955&1.333&7.236&9.648&77.588&135.005\\
27.85&9.212&1.349&7.240&4.943&75.483&122.982\\
28.85&9.466&1.363&7.240&$5.899$&73.592&119.295\\
30.85&9.969&1.381&7.235&$-$&70.168&114.425\\
35.85&11.176&1.417&7.194&$-$&64.144&105.945\\
40.85&12.333&1.433&7.130&$-$&59.935&100.169\\
46.85&13.669&1.437&7.055&$-$&56.349&95.361\\

\hline
\end{tabular}

\end{center}

\label{eos1}

\end{table}

\newpage

\begin{table}[h]

\caption{Data for EOS3 (SS2)}

\vskip 1cm

\begin{center}

\begin{tabular}{|c|c|c|c|c|c|c|}

\hline

$\rho_c \times~10^{14}$&$n_c/n_0$&$M/M_{\odot}$&R&$\omega_0 \times
~10^3$&$\omega_1 \times~10^3$&$\omega_2 \times ~10^3$\\
g/c.c.&&&km&/sec.&/sec.&/sec.\\

\hline

17.17&6.067&0.423&5.070&86.879&195.416&299.090\\
18.17&6.382&0.539&5.460&74.016&172.966&264.894\\
19.17&6.695&0.659&5.794&63.013&153.699&236.761\\
20.17&7.006&0.781&6.078&53.079&136.745&212.463\\
21.17&7.298&0.855&6.227&47.332&127.643&199.063\\
22.17&7.588&0.923&6.351&42.074&119.663&186.932\\
23.17&7.876&0.986&6.453&37.131&112.402&176.070\\
24.17&8.156&1.036&6.524&33.069&106.694&167.790\\
25.17&8.428&1.075&6.575&29.646&102.152&161.274\\
26.17&8.699&1.110&6.615&26.321&98.030&155.270\\
27.17&8.967&1.142&6.647&22.992&94.193&149.689\\
28.17&9.227&1.167&6.667&20.163&91.206&145.298\\
29.17&9.485&1.188&6.682&17.335&88.518&141.539\\
30.17&9.741&1.207&6.693&14.307&86.006&137.932\\
31.17&9.994&1.224&6.691&10.827&83.642&134.554\\
32.17&10.242&1.237&6.702&6.982&81.741&131.825\\
33.17&10.488&1.249&6.703&$3.351$&79.959&129.293\\
35.17&10.974&1.270&6.698&$-$&76.694&124.666\\
40.17&12.148&1.301&6.650&$-$&70.680&116.215\\
45.17&13.278&1.316&6.622&$-$&66.380&110.292\\
50.17&14.371&1.323&6.573&$-$&63.141&105.918\\
55.17&15.537&1.325&6.518&$-$&60.616&102.546\\

\hline
\end{tabular}

\end{center}

\label{eos3}

\end{table}

\newpage

\begin{table}[htbp]

\caption{Data for SS1 from fitted EOS1}

\vskip 1cm

\begin{center}

\begin{tabular}{|c|c|c|c|c|c|c|}

\hline

$\rho_c \times ~10^{14}$&$n_c/n_0$&$M/M_{\odot}$&R&$\omega_0
\times~10^3$&$\omega_1 \times~10^3$&$\omega_2 \times ~10^3$\\
g/c.c.&&&km&/sec.&/sec.&/sec.\\

\hline

14.85&5.436&0.635&6.188&59.977&144.353&222.426\\
15.85&5.763&0.770&6.544&49.791&127.744&196.858\\
16.85&6.081&0.879&6.787&42.407&114.994&179.262\\
17.85&6.392&0.968&6.957&36.671&106.205&166.319\\
18.85&6.697&1.041&7.079&31.962&99.289&156.335\\
19.85&6.994&1.101&7.167&27.938&93.826&148.390\\
20.85&7.287&1.152&7.230&24.376&89.333&141.887\\
21.85&7.573&1.194&7.276&21.115&85.580&136.461\\
22.85&7.855&1.229&7.307&18.025&82.380&131.844\\
23.85&8.132&1.260&7.329&14.976&79.615&127.894\\
24.85&8.405&1.286&7.342&11.780&77.195&124.449\\
25.85&8.674&1.308&7.350&8.036&75.052&121.420\\
26.85&8.939&1.327&7.352&1.034&73.175&118.733\\
27.85&9.201&1.343&7.350&$7.078$&71.465&116.329\\
30.85&9.840&1.380&7.329&$-$&67.274&110.448\\
35.85&11.184&1.415&7.265&$-$&62.242&103.549\\
40.85&12.343&1.431&7.187&$-$&58.683&98.749\\
46.85&13.671&1.436&7.091&$-$&55.523&94.636\\

\hline
\end{tabular}

\end{center}

\label{ss1}

\end{table}

\newpage

\begin{table}[htbp]

\caption{Data for SS2 from fitted EOS3}

\vskip 1cm

\begin{center}

\begin{tabular}{|c|c|c|c|c|c|c|}

\hline

$\rho_c~ \times 10^{14}$&$n_c/n_0$&$M/M_{\odot}$&R&$\omega_0
\times ~10^3$&$\omega_1 \times ~10^3$&$\omega_2 \times ~10^3$\\
g/c.c.&&&km&/sec.&/sec.&/sec.\\

\hline

17.17&6.036&0.579&5.718&64.793&155.655&239.799\\
18.17&6.351&0.688&6.013&55.081&139.070&215.380\\
19.17&6.659&0.779&6.224&47.803&127.136&197.858\\
20.17&6.961&0.855&6.378&42.037&118.072&184.600\\
21.17&7.257&0.919&6.494&37.269&110.906&174.149\\
22.17&7.547&0.973&6.581&33.177&105.095&165.697\\
23.17&7.833&1.019&6.646&29.567&100.382&158.683\\
24.17&8.114&1.059&6.696&26.300&96.185&152.771\\
25.17&8.390&1.093&6.733&23.270&92.680&147.719\\
26.17&8.663&1.122&6.761&20.392&89.632&143.331\\
27.17&8.931&1.148&6.781&17.584&86.947&139.496\\
28.17&9.196&1.170&6.795&14.749&84.584&136.101\\
29.17&9.458&1.190&6.804&11.724&82.467&133.077\\
30.17&9.716&1.207&6.809&8.129&80.545&130.365\\
31.17&9.971&1.222&6.811&1.818&78.823&127.914\\
32.17&10.224&1.235&6.809&$6.940$&77.243&125.684\\
35.17&10.964&1.266&6.795&$-$&73.257&120.094\\
40.17&12.140&1.298&6.747&$-$&68.293&113.270\\
45.17&13.286&1.314&6.688&$-$&64.668&108.365\\
50.17&14.379&1.322&6.624&$-$&61.862&104.656\\
55.17&15.332&1.324&6.561&$-$&59.601&101.747\\ \hline
\end{tabular}
\end{center}
\label{ss2}
\end{table}

\begin{table}[htbp]
\caption{Data for bag model with B=60 \& ms=150} \vskip 1cm
\begin{center}
\begin{tabular}{|c|c|c|c|c|c|c|}
\hline

$\rho_c \times ~10^{14}$&$n_c/n_0$&$M/M_{\odot}$&R&$\omega_0
\times ~10^3$&$\omega_1 \times ~10^3$&$\omega_2 \times ~10^3$\\
g/c.c.&&&km&/sec.&/sec.&/sec.\\

\hline

6.20&2.421&0.691&8.549&38.964&92.549&142.426\\
7.20&2.778&1.019&9.544&27.409&73.557&114.644\\
8.20&3.122&1.240&10.021&20.733&63.870&100.636\\
9.20&3.454&1.393&10.263&15.915&57.854&92.024\\
10.20&3.776&1.501&10.393&11.910&53.693&86.161\\
11.20&4.089&1.581&10.452&7.900&50.628&81.883\\
12.20&4.396&1.639&10.469&2.159&48.272&78.626\\
13.20&4.695&1.683&10.462&$6.156$&46.395&76.048\\
15.20&5.277&1.741&10.405&$-$&43.526&72.203\\
17.20&5.839&1.775&10.321&$-$&41.429&69.514\\
23.70&7.560&1.805&10.012&$-$&37.316&64.404\\

\hline
\end{tabular}

\end{center}

\label{bag1}

\end{table}

\begin{table}[htbp]
\caption{Data for bag model with B=75 \& ms=150} \vskip 1cm
\begin{center}
\begin{tabular}{|c|c|c|c|c|c|c|}
\hline

$\rho_c \times ~10^{14}$&$n_c/n_0$&$M/M_{\odot}$&R&$\omega_0
\times  ~10^3$&$\omega_1 \times ~10^3$&$\omega_2 \times ~10^3$\\
g/c.c.&&&km&/sec.&/sec.&/sec.\\

\hline

9.83&3.573&1.072&8.923&24.809&73.494&115.482\\
10.83&3.892&1.198&9.148&20.030&67.175&106.403\\
11.83&4.203&1.293&9.281&16.097&62.623&99.944\\
12.83&4.506&1.366&9.356&12.565&59.194&95.089\\
13.83&4.804&1.422&9.396&9.063&56.480&91.314\\
14.83&5.095&1.467&9.412&4.502&54.280&88.272\\
15.83&5.318&1.502&9.411&$4.887$&52.470&85.769\\
20.83&6.748&1.594&9.294&$-$&46.505&77.834\\
25.83&8.031&1.622&9.123&$-$&43.109&73.610\\
28.83&8.769&1.626&9.020&$-$&41.638&71.868\\

\hline
\end{tabular}

\end{center}

\label{bag2}

\end{table}


\begin{thebibliography}{}
\bibitem{d98} M. Dey, I. Bombaci, J. Dey, S. Ray  \& B. C.
Samanta, Phys.  Lett. {\bf B438} 123 (1998) ; Addendum  B447
(1999) 352; Erratum B467 (1999) 303;  Indian J. Phys. 73B (1999)
377.

\bibitem{jjd02} J. J. Drake et al, Is RXJ1856.5$-$3754 a Quark Star?,
astro-ph/0204159 v1, Ap. J. {\bf 572} (2002) 996.

\bibitem{anderson} D. P. Anderson, E. T. Herrin, V. L. Teplitz and I. M. Tibulaec,
`Two Seismic Events with the Properties for the Passage of Strange Quark Matter Through the Earth', Astro-ph/0205089.

\bibitem{Li99a} X. Li, I. Bombaci, M. Dey, J. Dey  \& E. P. J. van den
Heuvel, Phys. Rev. Lett. {\bf 83} 3776 (1999).

\bibitem{Li99b} X. Li, S. Ray, J. Dey, M. Dey  \& I.  Bombaci,
Ap. J. {\bf 527} L51 (1999).

\bibitem{fr01} L. M. Franco, The Effect of Mass Accretion Rate on
the Burst Oscillations in 4U~1728-34. astro-ph/0009189, Ap. J.
{\bf 554} 340 (2001).

\bibitem{bbb97aa}  M. Baldo, I. Bombaci  \&  G. F. Burgio,
Astron. \& Astrophys. {\bf 328} 274 (1997).

\bibitem{sinha} M. Sinha, M. Dey, S. Ray and J. Dey, Superbursts
and long bursts as surface phenomenon of compact objects (to be
published).

\bibitem{gbzgrdd00aa}  D. Gondek-Rosi\'nska , T.  Bulik , L.  Zdunik ,
E. Gourgoulhon , S. Ray , J. Dey  \& M. Dey, Astron. \& Astrophys.
{\bf 363} 1005 (2000).

\bibitem{qnova} R. Ouyed, J. Dey and M. Dey, astro-ph/0105109v3,
Astron. \& Astrophys. (in press).

\bibitem{wit} E. Witten, Phys. Rev. D {\bf 30} 272 (1984).

\bibitem{r2000} S. Ray, J. Dey , M. Dey , K. Ray  \&  B. C.
Samanta,  Astron. \& Astrophys. {\bf 364} L89 (2000).

\bibitem{p78n} J. V. Paradijs, Nature {\bf 274} 650 (1978); J. V. Paradijs,
Ap. J. {\bf 234} 609 (1979); J. V. Paradijs, Astron. \& Astrophys.
{\bf 101} 174 (1981).

\bibitem{a91npb} C. Alcock, Nucl. Phys. B. (Proc. Suppl.) {\bf 24} 93 (1991).

\bibitem{u1} V. V. Usov, Ap. J. 481 (1997) L107. ; V. V. Usov, Phys. Rev.
Lett. {\bf 80} 230 (1998).

\bibitem{m98} A. Mitra, Phys. Rev. Lett., {\bf 81} 4774 (1998).

\bibitem{u3} V. V. Usov, Phys. Rev. Lett. {\bf 81} 4775 (1998).

\bibitem{u4} V. V. Usov, Ap. J.  {\bf 550} L179 (2001).

\bibitem{u5} V. V. Usov, Phys. Rev. Lett.{\bf 87} 021101 (2001).

\bibitem{xqz99apj} R-X. Xu,  G. J. Qiao  \& B. Zhang,  Ap. J.
{\bf 522} L109 (1999).

\bibitem{ks} R. C. Kapoor  and  C. S. Shukre, ``Are radio pulsars strange
stars?" astro-ph/0011386, Astron. and Astrophys. {\bf 375} 405
(2001).

\bibitem{xxw} R-X. Xu, X-B. Xu \& X-J. Wu, Chin. Phys. Lett. {\bf
18} 837 (2001).

\bibitem{srdd00mpla} S. Ray, J. Dey  \& M. Dey  Mod. Phys. Lett.
A {\bf 15} 1301 (2000).

\bibitem{g00prl} N. Glendenning,  Phys. Rev. Lett. {\bf 85} 1150 (2000).

\bibitem{sharma} R. Sharma, S. Mukherjee, Mira Dey and Jishnu Dey,
Mod. Phys. Lett. A {\bf 17} 827 (2002).

\bibitem{cdwl98s} K. S. Cheng, Z, G. Dai, D. M. Wai  \& T. Lu
Science, {\bf 280} 407 (1998).

\bibitem{cd98prl} K. S. Cheng  \&  Z. G. Dai, Phys. Rev.
Lett. {\bf 80} 18 (1998) .

\bibitem{boma} I. Bombaci and B. Datta, Ap. J. {\bf 530} L69 (2000).

\bibitem{muno} M. P. Muno, D. Chakrabarty, D. K. Galloway and P.
Savov, Ap. J. {\bf 553} L157 (2001).

\bibitem{chandra} R. Wijnands, J. M. Miller, C. Markwardt, W. H.
G. Lewin and M. van der Klis, `A {\it Chandra} observation of the
long-duration X-ray transient KS~1731$-$260 in quiescence : too
cold a neutron star?', astro-ph/0107380, submitted to Ap. J.
Letters.

\bibitem{chan} S. Chandrasekhar, Phys. Rev. Lett. {\bf 12} 114 (1964) ; Ap. J, {\bf 217} 417 (1964) .

\bibitem{kok} K.D. Kokkotus and J. Ruoff, A \& A A, {\bf 366} 565 (2001) ; V.K.
Gupta, Vinita Tuli and Ashoke Goyal, astro-ph/0202016.

\bibitem{mis}C.W. Misner, K.S. Throne \& J. A. Wheeler, {\it Gravitation},
W. H. Free man \& Co. New York (1973).

\bibitem{pons}Jose A. Pons et al , Ap. J., {\bf 564} 981 (2002).

\bibitem{pr} A.R. Prasanna and Subharthi Ray, astro-ph/0205343.

\end{thebibliography}
\end{document}